\documentstyle [12pt]{article}

\begin{document}
\centerline{QUANTUM CONDUCTANCE FLUCTUATIONS}
\centerline{IN 3D BALLISTIC ADIABATIC WIRES}

$$ $$

\centerline{Vladimir I. Fal'ko$^{a,b)}$ and G.B. Lesovik$^{b)}$}
\centerline{$^{a)}$ {\it Max-Planck-Institut f\"ur Festk\"orperforschung,
Stuttgart, Germany\/}}
\centerline{$^{b)}$ {\it Institute of Solid State Physics, Chernogolovka,
142432 Russia\/}}

$$ $$
$$ $$

During recent years the transport properties of ballistic adiabatic
conductors were extensively studied.  The most of the interest in
this field has been attracted by the 2D semiconductor devices, so that
the features of a quantum transport in a ballistic adiabatic wire
prepared of a bulk metal or semimetal has not been discussed.  In
the present paper we study the quantum conductance of three
dimensional ballistic wires with idealy flat boundaries and show that
it obeys fluctuations with the properties quite distinguishable from
those of the universal conductance fluctuations (UCF, \cite{UCF}):
both the fluctuations amplitude and the sensitivity of the conductance
to the magnetic field flux $\Phi =HS$ penetrated into the sample
cross-sectional area $S$ are different and depend on details of the
shape of a wire.  When the wire has the cross section with the
shape of an integrable billiard, conductance fluctuations have the
enlarged amplitude $\delta G\sim\left[(e^2/h)^3G\right]^{1/4}$
and the universal correlation
magnetic field $H_c\sim\Phi_0/S$.  When the cross-sectional shape of a wire
is non-integrable, the irregular part of a conductance has the $e^
2/h$
scale whereas the correlation field is reduced to the value of
$H_S\sim (\lambda_F/\sqrt S)^{1/2}(\Phi_0/S)$ and the correlation voltage
of the nonlinear
conductance fluctuations has the scale of $eV_c\sim\hbar^2/mS\sim
E_F/(S/\lambda_F)$, where $\lambda_F=1/p_F$ is the Fermi wavelength.

The following analysis is based on the application of the
Landauer-Buttiker approach.  That is, the two-terminal conductance $
G$
can be written as $G={{e^2}\over h}Tr(\hat {t}\hat {t}^{+})$, where $
\hat {t}$ is the scattering matrix
\cite{Buttiker}.  In the limit of a ballistic transport in long wires,
$t_{nm}=\delta_{nm}\times (0\:{\rm o}{\rm r}\:1)$ and the separation of
variables of the electron
motion along and across wire axis reduces the conductance formula to
the form \cite{Wees}
$$G={{e^2}\over h}N\left(E_F\right)$$
which includes now the number of reflectionless
quasi-one-dimensional channels $N(E_F)$.  The latter quantity is equal
to the number of size-quantized energy levels in a 2D box (with the
shape of a wire cross section) below the Fermi level in the system.
The main contribution to $N(E_F)$ is determined by the average density
of states and depends only on the area $S$ independently of the shape:
$N_0(E_F)\sim E_F/(2\pi \hbar^2/2mS)$.  Next, one can distinguish the term
which
manifests the features of the boundary conditions of the electron
wave function along the surcomference $L\sim\sqrt S$ of the wire surface.
When those are $\psi =0$, a strip $L\times\lambda_F$ should be excluded
from the
cross-sectional area of a sample and, which reduces $N_0$ by the
number $N_1=\theta\left(E_F/(\hbar^2/2mL^2)\right)^{1/2}$ \cite{Weyl}.
These two terms give
us the waveguide analogue to the Sharvin resistance formula
\cite{Sharvin}.  Finally, the exact number of states $N(E_F)$ depends on
the Fermi energy irregularly, and its deviation, $\delta N(E_F)$, from the
best smoothed approximation gives a subject for a statistical
consideration.  Altogether, these three contributions can be combined
into the conductance

\begin{equation}
G=\langle G\rangle +\delta G={{e^2}\over h}\left[{S\over {2\pi \lambda_
F^2}}-{L\over {\lambda_F}}\right]+{{e^2}\over h}\delta N\left(E_F\right
).
\label{fluctG}
\end{equation}
A randomly varying part of a conductance, $\delta G$ will be subject of
the following analysis.  In Ref.  \cite{We} these fluctuations have
been called {\bf geometrical conductance fluctuations} in order to
distinguish them from the UCF \cite{UCF}.

As we know from the quantum billiards theory, the statistics of
fluctuations $\delta N(E_F)$ depend on the integrability of the particle
motion
in a quantum box.  According to \cite{Bohigas,Berry}, the fluctuations
of the number of states below the Fermi level are the strongest in
integrable quantum systems:  the systems which possess an
additional integral of motion besides the energy.  The following
reasoning gives a handwaving estimation of this effect.  The
spectrum $\epsilon (I,n)$ of a particle confined in the integrable box can
be
imagined as a set of independent level series marked by different
values of a quantum number $I$, so that at energies $E\gg h^2/mS$ higher
than the mean level spacing the total spectrum is locally composed
of uncorrelated contributions from different rigid staircases of levels
derived from separate one-dimensional Hamiltonians (each
corresponding to some value of $I$).  This means that at the short
energy range $\epsilon$ the spectra of integrable systems obey the Poisson
type of statistics, $\langle (\delta [N(E+\epsilon )-N(E)])^2\rangle$ $
\approx{1\over {15}}\langle N(E+\epsilon )-N(E)\rangle$
\cite{Berry}, while the width of a spectral interval $\epsilon$ is small
enough
to consider all levels $\epsilon (I,n)$ inside as taken independently from
different series $I$.  Otherwise, the mean square $\langle (\delta
[N(E+\epsilon )-N(E)])^2\rangle$
is restricted by the number $I_{max}$ of independent level series
contributing to the spectrum formation, since each of them has a
strong internal rigidity.  In the case of a free particle moving in a
2D box, $\epsilon (I,n)\propto (I^2,n^2)$, so that the above-mentioned
Poisson law is
applicable only if $\epsilon <\epsilon_{max}\sim E_F/\sqrt {S/\lambda^
2_F}$.  Beyond this scale, the
amplitude of spectral fluctuations is saturated, what can be estimated
as $\langle (\delta [N(E_F+\epsilon )-N(E_F)])^2\rangle$ $=\langle
\delta N(E_F)^2\rangle\sim N(E_F)^{1/2}$.  It is amusing to
note that the problem of calculation of $\delta N(E_F)$ in rectangular
billiards is familiar to the number-theory as a problem of an
accuracy of the best smoothed series approximation of a number of
square lattice vortices inside an ellipse.  One of the best
number-theory results \cite{Rem} has predicted $\left\langle \delta
N^2\right\rangle \propto\left\langle N(E)\right\rangle ^{\theta}$,
where $0.6416\ge\theta\ge 1/2$, which agrees with the above qualitative
reasoning.

All this gives an estimation of the mean square value of the
conductance fluctuations $\delta G$ in the wire with an integrable cross
sectional shape,

\begin{equation}
\left\langle \delta G^2\right\rangle =\beta\alpha\left[\left({{
e^2}\over h}\right)^3G\right]^{{1\over 2}}.
\label{Int1} \end{equation}

The scale of these fluctuations can exceed the quantum ${{e^2}\over
h}$.
Here, $\alpha$ is a specific geometrical factor (for instance, $\alpha
=0.095$ in a
rectangle with practically equal sides) and $\beta$ accounts for the
symmetry induced degeneracy of states.

The spectra of non-integrable systems are much more rigid, and the
fluctuations in them are rather weak.  In terms of the interaction
of levels, this results from the repulsion between all of them
(not only within each spectral series, as in the case of integrable
billiards).  The limiting case of shuc levels statistics is given by that
of chaotic billiards \cite{Berry}, i.e.,  the quantity $\delta N\left
(E_F\right)$ obeys some
kind of a saturated Wigner-Dyson law \cite{Wigner}.  Therefore, the
amplitude of geometrical conductance fluctuations in wires of an
arbitrary shape can be estimated as

\begin{equation}
\left\langle \delta G^2\right\rangle =\beta\left({{e^2}\over {\pi
h}}\right)^2\ln\left({\rm l}{\rm n}{{Gh}\over {e^2}}\right).
\label{NonInt1}
\end{equation}

{\bf Fluctuations of nonlinear conductance.}  Since neither the shape nor
the
Fermi level can be easily varied in 3D metallic wires, the most
natural possibility to observe the spectral fluctuations of the
transverse motion in an adiabatic wire consists in the studies of its
diferential (nolinear) conductance $dI(V)/dV$.  In the adiabatic regime,
this quantity can be expressed in terms of a number of transmitted
waveguide modes \cite{Glazman} as

$${{dI(V)}\over {dV}}={{e^2}\over {2h}}\left[N\left(E_F+{{eV}\over
2}\right)+N\left(E_F-{{eV}\over 2}\right)\right].$$

One can see from this equation that the voltage dependence of
$dI(V)/dV$ just follows the local (in energy) fluctuations of the
number of size-quantized states in the interval $eV$ near the Fermi
level.  In a constriction with the cross section presenting completely
integrable billiard, this value undergoes fluctuations with the Poisson
statistics at low voltages $V<V_c=\,\epsilon_{max}/e$.  When $V\ge
V_c$, the
amplitude of fluctuations saturates at the value described by Eq.
(\ref{Int1}) and then $V_c$ plays the role of the correlation voltage of
these fluctuations.  Therefore, in the adaibatic contact with an
integrable cross-sectional shape

\begin{equation}
\left\langle \left[{{dI(V)}\over {dV}}-G\right]^2\right\rangle =\left
\{\matrix{{1\over {15}}\left({{eV}\over {E_F}}\right){{e^2}\over h}
G,\hbox{    $eV<\epsilon_{max}$ }\cr
\cr
\gamma\sqrt {\left({{e^2}\over h}\right)^3G}\hbox{,    }eV>\epsilon_{
max}\cr}
\right..
\label{diffG1}
\end{equation}

In {\it non-integrable\/} (i.e.,  chaotic) systems the spectral
fluctuations are
weaker \cite{Berry,Wigner},

\begin{equation}
\left\langle \left[{{dI(V)}\over {dV}}-G\right]^2\right\rangle =
3\beta\left({{e^2}\over {\pi h}}\right)^2\left\{\matrix{\ln\left({{
VGh}\over {eE_F}}\right)\hbox{,  }eV<\epsilon^{\prime}_{max}\cr
\cr
\ln\left(\ln\left({{Gh}\over {e^2}}\right)\right)\hbox{,  }eV>\epsilon^{
\prime}_{max}\cr}
\right..
\label{diffG2}
\end{equation}
The correlation voltage which one could assign to these fluctuations
is determined by the mean level spacing $\hbar^2/mS$ and has the form
\begin{equation}
V_c={1\over e}{{E_F}\over {S/\lambda_F^2}}.
\label{Vc}\end{equation}
It is interesting to note that this result differs from the
time-of-flight estimation which one could get after replacing the
mean free path by the sample length in the formulae valid for
diffusive conductors  \cite{Larkin}.

{\bf Magnetoconductance fluctuations.}  Another possibility to observe the
geometrical conductance fluctuation consists in an application of a
magnetric field oriented along the axis of a wire.  In what follows,
we distinguish three cases of different cross-sectional shapes of a
wire.

Let us first consider the  {\it system which is integrable and retains
this property also in an applied magnetic field\/} (the disk obviously
belongs to this class).  The external magnetic field shifts series of
levels marked by different value of the angular momentum
$I$, one with respect to another.  When this shift,
${{\hbar e}\over {mc}}HI\sim H{{\hbar e}\over {mc}}\sqrt {E_F/(\hbar^
2/2mS)}$, becomes comparable to the mean
intra-series level spacing, $\sqrt {E_F\hbar^2/2mS}$, the statistical
configuration of
the Poisson distribution of levels is renewed, which produces a
random magnetoconductance variation of the order of what is
represented in Eq.  (\ref{Int1}) ($\beta =1$).  The characteristic scale of
a
sufficient magnetic field is determined by the flux quantum
penetrated through the sample cross-sectional area, $H_cS\sim hc/e$,
which means that the important physical quantity - the correlation
magnetic field of fluctuations \cite{UCF} - is
similar to that in UCF.  The visual difference between UCF and
geometrical fluctuations is only in their amplitudes.  The geometrical
fluctuations are enlarged parametrically in thick cylindrical wires,
what has been observed in \cite{Bogachek} in numerical simulations.
On the other hand, these fluctuations can be viewed (even inside the
correlation field $H_c)$ as a series of randomly distributed $e^2/
h$
conductance steps with the characteristic spacing of
$\Phi_S\sim{{\Phi_0/\sqrt {\alpha}}\over {^4\sqrt {N(E_F)}}}\sim\Phi_
0\left[{{2e^2/h}\over {\alpha^2G}}\right]^{{1\over 4}}$ .

Of course, one should realize that the wires with the cross-sectional
shape which is integrable at any magnetic field gives us an exclusive
example.  {\it Magnetoconductance fluctuations in wires with a
non-integrable cross-sectional shape \/} (see Eq.  (\ref{NonInt1})) have a
much lower amplitude what makes them more similar to the UCF.
Nevertheless, they still possess one feature specific to the ballistic
adiabatic system.  That is, the non-integrable quantum box spectrum
is much more sensitive to the magnetic field variation that it takes
place in open diffusive conductors, and the penetration of a flux
\begin{equation}
\Phi_S=\sqrt {{{\lambda_F}\over {\sqrt S}}}{{hc}\over e}\ll\Phi_
0
\label{flux}\end{equation}
is already enough for renewing the realization of the levels
configuration \cite{Stone}.  In a chaotic billard, all the classical
electron trajectories are infinitely long and cover the whole
fixed-energy thorus in the phase space.  In semiclassics, the length
of such trajectories is limited, since two point of the phase space
which reach into the same unit volume $dp_idx_i\sim h$ are
indistinguishable.  Therefore, the length of a semiclassical trajectory
in a box of characteristic dimensions $L\sim\sqrt S$ should be cut when the
inverse time of flight along it, $h\sqrt S/v_F$ will be comparable with the
mean level spacing $\Delta\sim\hbar^2/mS$ and characteristic traces in the
real
space contain $\sim\sqrt S/\lambda_F$ closed loops each encircling an area
$
\sim S$ of a
random sign.  The oriented encircled area of a chaotic trajectory can
be estimated as $S^{3/2}/\lambda_F$ which produces the characteristic
correlation magnetic field $H_c\sim\Phi_S/S$.  It would be interesting to
note in this connection that the features of the conductance
fluctuations found in ballistic silver micro-contacts \cite{Exp} are
quite similar to those of wires with a non-itegrable shape described
by Eqs.  (\ref{Vc},\ref{flux}), though the observed amplitudes were
much smaller than $e^2/h$ scale.

Finally, there exists a class of {\it systems which are integrable\/} (or
partly integrable \cite{Berry}) {\it at $H=0$ and loose this property after
an application of a magnetic filed\/}.  Fot instance, we can mention the
structures with a rectangular cross-sectional shape where the first
penetrated flux quantum drastically changes the level statistics:
from the Poisson type at $\Phi =HS<\Phi_0$ to the Wigner-Dyson type at
higher fields.  This should induce a low-field magnetoconductance of a
random sign with the amplitude estimated by Eq.  (\ref{Int1}),
whereas the following increase of a produces fluctuations with the
amplitude ${{e^2}\over h}$ and at the scale of $\Phi_S$ (instead of $
\Phi_0$ observed in the
UCF).  The monocrystalline microwires with perfect facets (whiskers)
grown of the semimetallic material would be the best candidates to
show this kind of behavior.

We are greateful to  P. Wyder and A.  Jansen for helpful discussions.

\end{document}